\documentclass{jetpl}
\addtocounter{table}{-1}
\usepackage{graphicx}
\twocolumn

\lat

\title{Optimized surface ion trap design for tight  confinement and separation of ion chains}

\rtitle{Optimized surface ion trap\ldots}

\sodtitle{Optimized surface ion trap design for tight  confinement and separation of ion chains}

\author{
I.\,S.\,Gerasin$^{1,2,3,}$\thanks{e-mail: i.gerasin@rqc.ru}, N.\,O.\,Zhadnov$^{1}$, K.\,S.\,Kudeyarov$^{1}$, K.\,Y.\,Khabarova$^{1}$, N.\,N.\,Kolachevsky$^{1,2}$, 
I.\,A.\,Semerikov$^{1,2}$
}

\rauthor{I.\,S.\,Gerasin, N.\,O.\,Zhadnov, K.\,S.\,Kudeyarov, K.\,Y.\,Khabarova, N.\,N.\,Kolachevsky,
I.\,A.\,Semerikov}

\sodauthor{Gerasin, Zhadnov, Kudeyarov, Khabarova, Kolachevsky, Semerikov}

\address{$^1$P.N. Lebedev Physical Institute, Russian Academy
of Sciences, Moscow, Russia; \\
$^2$Russian Quantum Center, Moscow, Russia \\
$^3$Moscow Institute of Physics and Technology,  Dolgoprudny, Russia \\}

\abstract{Qubit systems based on trapped ultracold ions win one of the leading positions in the quantum computing field, demonstrating quantum algorithms with the highest complexity to date. Surface Paul traps for ion confinement open the opportunity to scale quantum processors to hundreds of qubits and enable high-connectivity manipulations on ions. 
 To fabricate such a system with certain characteristics, the special design of a surface electrode structure is required. The depth of the trapping potential, the stability parameter, the secular frequency and the distance between an ion and the trap surface should be optimized for better performance.  Here we present the optimized design of a relatively simple surface trap that allows several important high-fidelity primitives: tight ion confinement, laser cooling, and wide optical access. The suggested trap design also allows to perform an important basic operation, namely, splitting an ion chain into two parts.}


\begin{document}

\maketitle


\section{Introduction}

The field of quantum computing is rapidly advancing. Using qubits as information carriers allows the implementation of new algorithms, which can overcome classical computing \cite{Grover1997, shor1994proceedings}. One of the possible ways of qubits realization is encoding the internal energy levels in atomic ions \cite{PhysRevLett.74.4091} which are confined by RF and DC electric fields in Paul traps \cite{PhysRevA.22.1137} and entangling ions via common vibrational modes in the trap. Ions are a promising platform for quantum computing due to their long lifetime in the trap, long coherence times of the qubit levels, straightforward initialization and readout, and strong Coloumb interaction between particles. 
Additionally, ultracold trapped ions are potential candidates for implementing quantum memristors \cite{spagnolo2022experimental}, which are promising elements for neuromorphic (biomimetic) computing systems. This is due to their numerous ion levels with varying lifetimes and transitions of different "oscillator strengths"\, as well as the ability of using full connectivity via vibrational modes \cite{stremoukhov2023proposal,stremoukhov2024model}.


One of the most pronounced challenges in advancing quantum computers lies in scaling them up, namely, increasing the number of individually controlled qubits while upholding quantum gate fidelities, low decoherence, and high connectivity \cite{bruzewicz2019trapped}. 3D Paul traps consisting of four massive RF electrodes allow to confine more than 100 ions in a linear chain useful for computation \cite{pagano2018cryogenic}. However, addressing individual ions in such lengthy chains poses a significant challenge due to the small interionic distances. Besides, maintaining the high fidelity of two-qubit gates begins to require an extremely high level of quantum state control due to the complication of the motional-mode spectrum of multi-ion chains \cite{leung2018entangling}.
To overcome these issues, the ion chain can be divided into smaller modules (sub-chains) by an external electric field. One can perform operations separately on such sub-chains and subsequently transfer quantum information between them by physically joining them up.
In 2002 the quantum charge-coupled device (QCCD) architecture was proposed \cite{kielpinski2002architecture}. In this approach, the ion trap is divided into several zones \cite{britton2009scalable} devoted to specific operation types, such as loading, initialization, quantum gates, storage, and readout. The advantage of such an architecture lies in its greater efficiency in utilizing experimental resources, such as lasers, by enabling the movement of ions into the interaction zone. This approach eliminates the need to create dedicated laser and optical systems for each ion.

To implement such an architecture, the concept of a linear Paul trap was converted to the form of a 2D microfabricated chip comprising a pattern of metal electrodes on the surface of a dielectric (quartz, silica, sapphire) or silicon substrate \cite{seidelin2006microfabricated, hughes2011microfabricated}. The manufacturing is based on well-established photolithographic techniques but also demands the use of the most suitable conducting and insulating materials \cite{romaszko2020engineering}, specific methods to increase the breakdown voltage \cite{sterling2013increased,wilson2022situ}, formation and pattering of thick dielectric layers, as well as minimizing the surface of dielectric observed by trapped ions due to patch charges.
The most impressive quantum computing performance to date has been shown by IonQ \cite{chen2023benchmarking} and Quantinuum \cite{moses2023race} companies, which demonstrated successful manipulation with 30 and 32 ion qubits in surface traps, respectively. Notably, Quantinuum achieved a remarkable quantum volume of $2^{16}$.

\begin{figure}[htp]
    \centering
            \includegraphics[width=3cm]{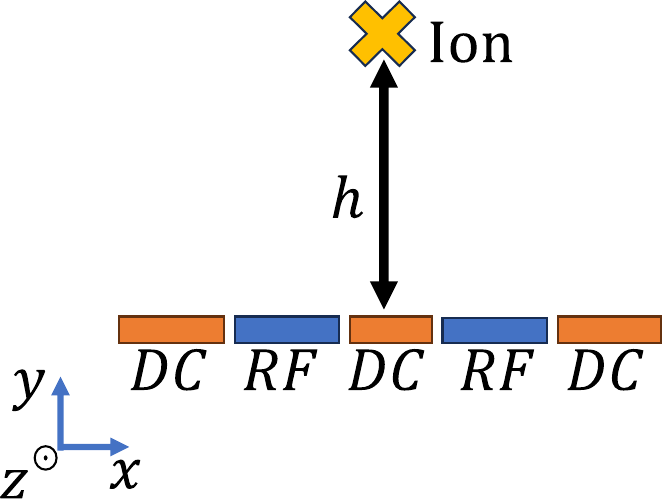}
            \includegraphics[width=3.5cm]{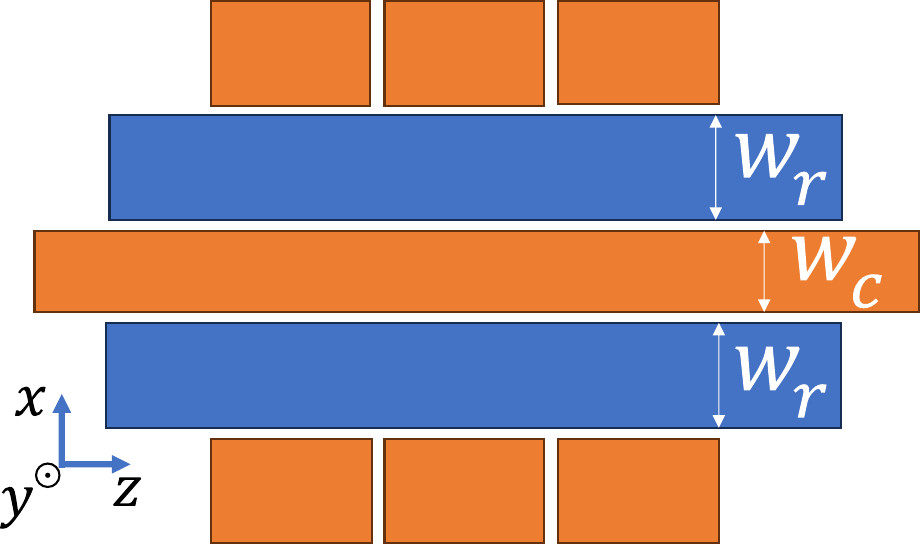}
        \caption{Fig. 1. Basic surface ion trap design: left - view along the trap axis, right -  top view.}
    \label{fig:trap}
\end{figure}

A basic design of a microfabricated surface trap is presented in Fig. \ref{fig:trap}. In this trap the radiofrequency (RF) voltage with an amplitude $V_{rf}$ and a frequency $f_{rf}$ is applied to a pair of electrodes. Between the RF electrodes, there is a grounded central electrode, forming the pseudopotential in the x-y plane that provides radial confinement of ions. Segmented outer electrodes are used to provide axial confinement (z-axis), relocate ions across the z-axis, and compensate micromotion \cite{berkeland1998minimization}.

\begin{figure}[htp]
    \centering
    \includegraphics[width=8.1 cm]{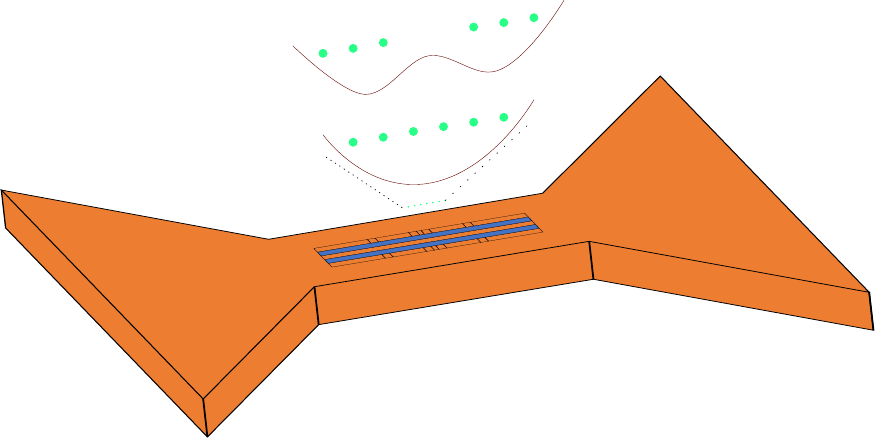}
    \caption{Fig. 2. Illustration of dividing a chain of 6 ions into two sub-chains. a - microchip with a surface ion trap, b - chain of 6 ions, c - two sub-chains after separation. The bow tie shape is standard for QCCD microchips, allowing wider optical access to the central ion trapping region.}
    \label{fig:chainsep3D}
\end{figure}

Surface traps make it relatively straightforward to perform the operations of separating and joining ion chains (Fig.\ref{fig:chainsep3D}). These operations are crucial for the realization of a flexible modular architecture, improving the fidelity of two-qubit gates, implementing two-qubit gates (such as physical SWAP) between ions in different chains, realizing error correction codes, and creating on-chip distributed quantum networks \cite{bruzewicz2019trapped}. To execute these operations, outer or center segmented DC electrodes can be used \cite{Nizamani2010}.

There is a number of studies focused on finding the optimal design for surface traps using both analytical models \cite{house2008analytic} and numerical simulations \cite{Nizamani2010, hong2016guidelines, abbasov2023surface}. In general, surface traps can consist of a large and complex system of electrodes, which are more easily described through computer simulations than by analytical methods. The main challenge in optimizing trap geometry lies in the numerous interconnected specifications that must be simultaneously optimized: trap depth, ion-to-trap distance, secular frequency, and stability parameter.

In this work, we investigate the influence of electrode size and geometry on the key trap parameters through simulations to identify a design that ensures robust confinement, efficient laser cooling and addressing, and the capability for ion chain separation.
Section 2 of this article describes the trap parameters and the calculation method. In Section 3, we optimize the electrode sizes for a basic surface trap. Sections 4 and 5 focus on designing a trap with asymmetric electrodes and a surface structure tailored for ion chain separation.
We hope the results of this study will serve as a valuable guide for designing surface ion traps.

\section{Methods}

Calculations were made for  $^{171}$Yb$^+$ as a qubit. There are several transitions in this ion to encode quantum information with long coherence times which are regularly used in quantum computing \cite{ryan2022implementing, wang2021single}, including our previous works \cite{aksenov2023realizing,zalivako2023continuous,zalivako2024towards,kazmina2024demonstration}.
Also it is worth note, that $^{171}$Yb$^+$ ion is extremely promising for the creation of quantum memristors \cite{stremoukhov2023proposal,stremoukhov2024model}.




To achieve better performance, the following trap parameters should be controlled: 
\begin{itemize}
\item  radial secular frequencies $f_{sec}$ characterizing the potential well:

\begin{equation}
    2 \pi f_{sec} = \frac{\epsilon Q V_{rf}}{\sqrt{2} m h^2 2 \pi f_{rf}}
\end{equation}

where $Q, m$ - the charge and the mass of the ion, $h$ - the distance to the nearest electrode (in our case the distance to the surface), $\epsilon$ - the efficiency parameter (typically 0.2 - 0.3 for surface traps) \cite{niedermayr2015cryogenic}. Usually, $f_{sec}$ is chosen to be about $2-4\,$MHz to provide the Lamb Dicke regime \cite{zalivako2024towards, mcloughlin2011versatile, kiesenhofer2023controlling}.
\item trap potential depth. It should be deep enough to capture particles produced by photoionization of the neutral atomic beam from the hot gun. The temperature of the atomic gun is about $600$ K which corresponds to particle energy of $\approx 0.05$ eV.
\item Mathieu stability parameter $q$:

\begin{equation}
    q = 2\sqrt2 \frac{f_{sec}}{f_{rf}} = \frac{2 \epsilon Q V_{rf}}{m h^2 (2 \pi f_{rf})^2},
    \label{eq:q}
\end{equation}

The stability parameter should be small enough ($q^2 \ll 1$) to maintain the harmonicity of ion oscillations. However, trap depth is proportional to this parameter \cite{leibfried2003quantum}.

\item the distance from the ion to the trap surface $h$ that defines the optical access required for laser cooling and quantum operations and impacts the ion heating rate induced by the surface ($\propto 1 / h^4$) \cite{brownnutt2015ion}. The optical access is determined by the numerical aperture (NA) for a beam propagating parallel to the trap plane and strongly focused on the ion. The NA is limited by the width of the isthmus of the trap (Fig.\ref{fig:chainsep3D}), which is typically around 1 mm. With an ion height above the trap surface ranging from $70$ to $100$ $\mu$m, the NA will be between 0.14 and 0.19 for a beam perpendicular to the isthmus. Such a setup will accommodate beam waists of $<2$ $\mu$m at the ion, which is sufficient for individual addressing.
\end{itemize}

The trap was simulated using a Python package Electrode \cite{electrode}, which allows one to evaluate the field distribution and pseudopotential parameters depending on the trap geometry. By searching through different configurations of electrodes, one can optimise the electrode structure for the desired confining potential.
The RF voltage amplitude is usually limited to several hundred volts due to electrical breakdown \cite{sterling2013increased,wilson2022situ}. We take a conservative estimation and set the amplitude $V_{rf} = 100\,$V.
We fix the $q$ parameter around 0.3, which is close to the optimal value, ensuring sufficient depth of the trapping potential. Since we use numerical estimations of the parameters, we can not always find a solution for exact value of $q$, so we search for it in the interval $q = 0.300 \pm 0.008$. If we want to consider higher voltage amplitudes while conserving $q$, we should increase the RF driving frequency $f_{rf}$ (Eq. \ref{eq:q}). These adjustments also result in the increase of $f_{sec}$, which is advantageous for the Lamb Dicke regime and the time of quantum gates. We assume that the frequency $f_{rf} \in [20, 24]\,$MHz which corresponds to $f_{sec} \approx 2\,$MHz.  By pre-setting the parameters ($V_{rf}, q, f_{sec}$), we manipulate the geometry of the symmetric surface trap to optimize the distance $h$ and the depth of the trapping potential. 


\section{Basic design}

We consider the planar trap configuration as depicted in  Fig.\ref{fig:trap}. The width of the central electrode is defined as  $w_c$, while the width of the RF electrodes is defined as $w_r$. We variate them in the following ranges: $w_c\in [30, 300]\,$$\mu$m, $w_r \in [30, 300]\,$$\mu$m. The outer electrodes are considered to be grounded and 1 mm wide. The gap between neighboring electrodes is set to 6 $\mu$m due to constraints associated with the fabrication process and the limitations related to the RF breakdown.  

The first step is to find the configuration of electrodes which optimizes the distance from the ion to the trap surface $h$. This parameter is solely determined by the trap geometry and is independent of voltage settings. In contrast, the secular frequencies, the stability parameter, and the potential depth are determined by both the amplitude and the frequency of the RF field. 

\begin{figure}[htp]
    \centering
    \includegraphics[width=7cm]{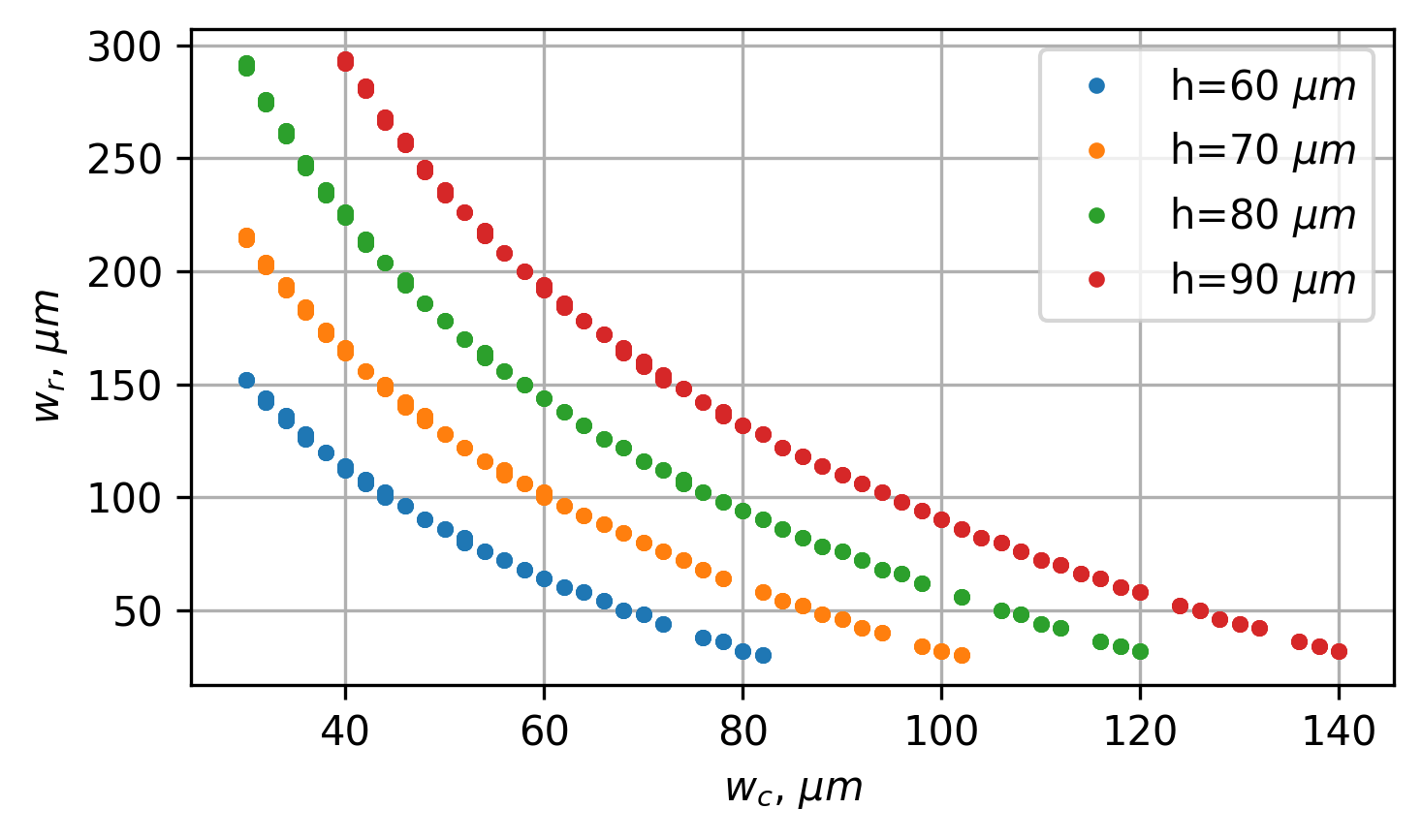}
    \caption{Fig. 3. Graphs show the dependence of the distance from the ion to the trap surface $h$ on the sizes of electrodes $w_r$ and $w_c$. 
    Parameters $q$ and $f_{sec}$ vary along the curves.}
    \label{fig:height}
\end{figure}

The dependence between the electrode's widths and the distance $h$ is presented in Fig. \ref{fig:height}. Notable, that increasing of $w_c$ and $w_r$ results in the increase of the distance $h$. It results from the linear scaling between the trap geometry and trapping potential. The points with fixed $h$ correspond to different potential depths, secular, and driving voltage frequencies. So, for the specified $h$, there exist several degrees of freedom, enabling adjustments in the widths of the electrodes.

\begin{figure}[htp]
    \centering
    \includegraphics[width=7cm]{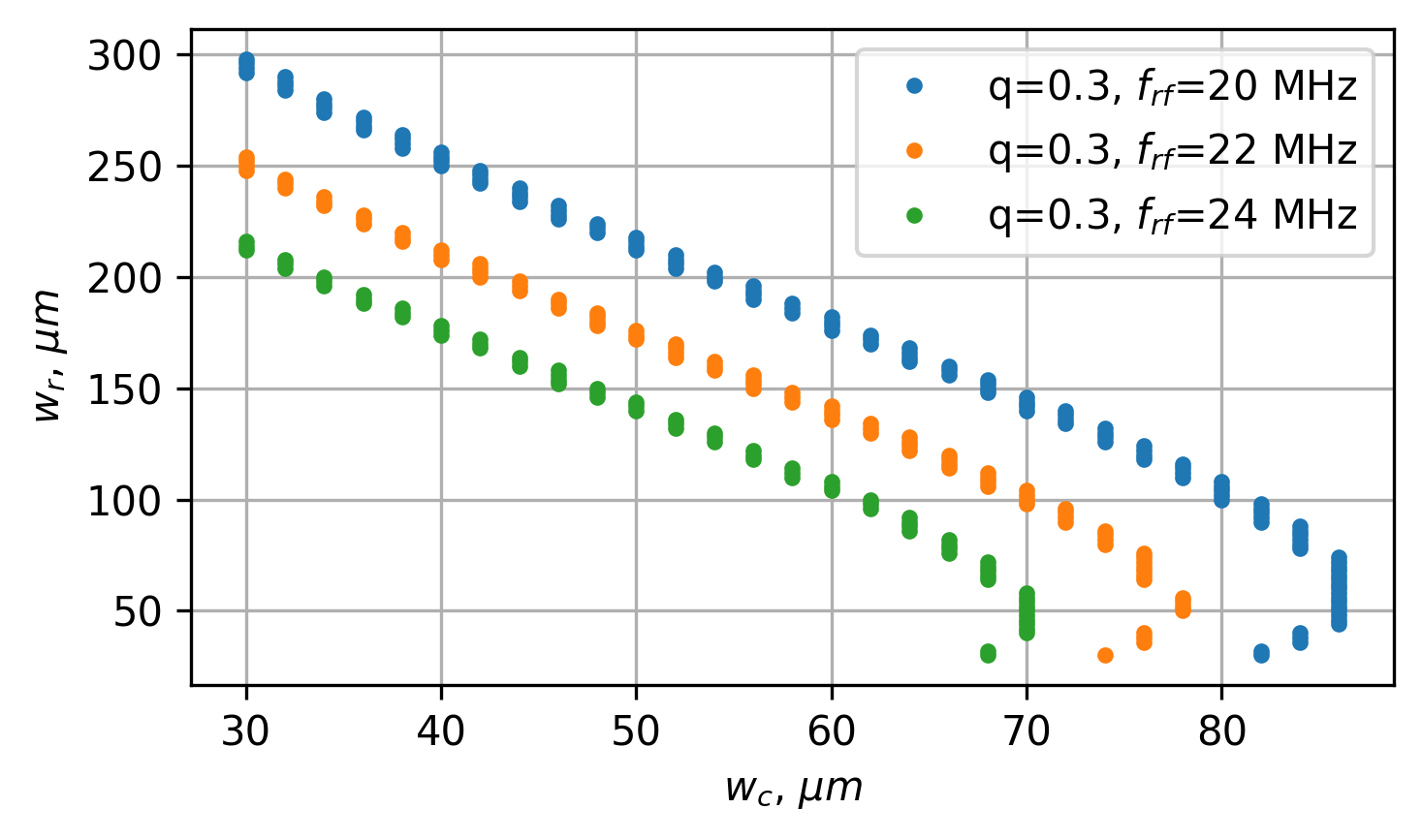}
    \caption{Fig. 4. Dependence of $w_r$ as a function of $w_c$ for fixed $q$ = 0.3, $V_{rf}$ = 100 V and three different values of $f_{rf}$.}
    \label{fig:freq}
\end{figure}

Fig. \ref{fig:freq} represents the dependencies between $w_c$ and $w_r$ for the case  when the stability parameter  $q$ and the driving frequency $f_{rf}$ (and therefore, the secular frequency $f_{sec}$) are fixed.



The plots presented in Fig. \ref{fig:height} and Fig. \ref{fig:freq} allow us to select a trap geometry satisfying the required parameters.
Fixing the distance $h$, the driving frequency $f_{rf}$, and the stability parameter $q$ defines a pair of curves. The intersection of these curves corresponds to the target geometry. Notably, there may be a combination of parameters for which the desired distance $h$ is not achievable because the curves do not intersect. Decreasing $h$, we obtain one or two solutions defining trap geometry. 

\begin{figure}[htp]
    \centering
    \includegraphics[width=8cm]{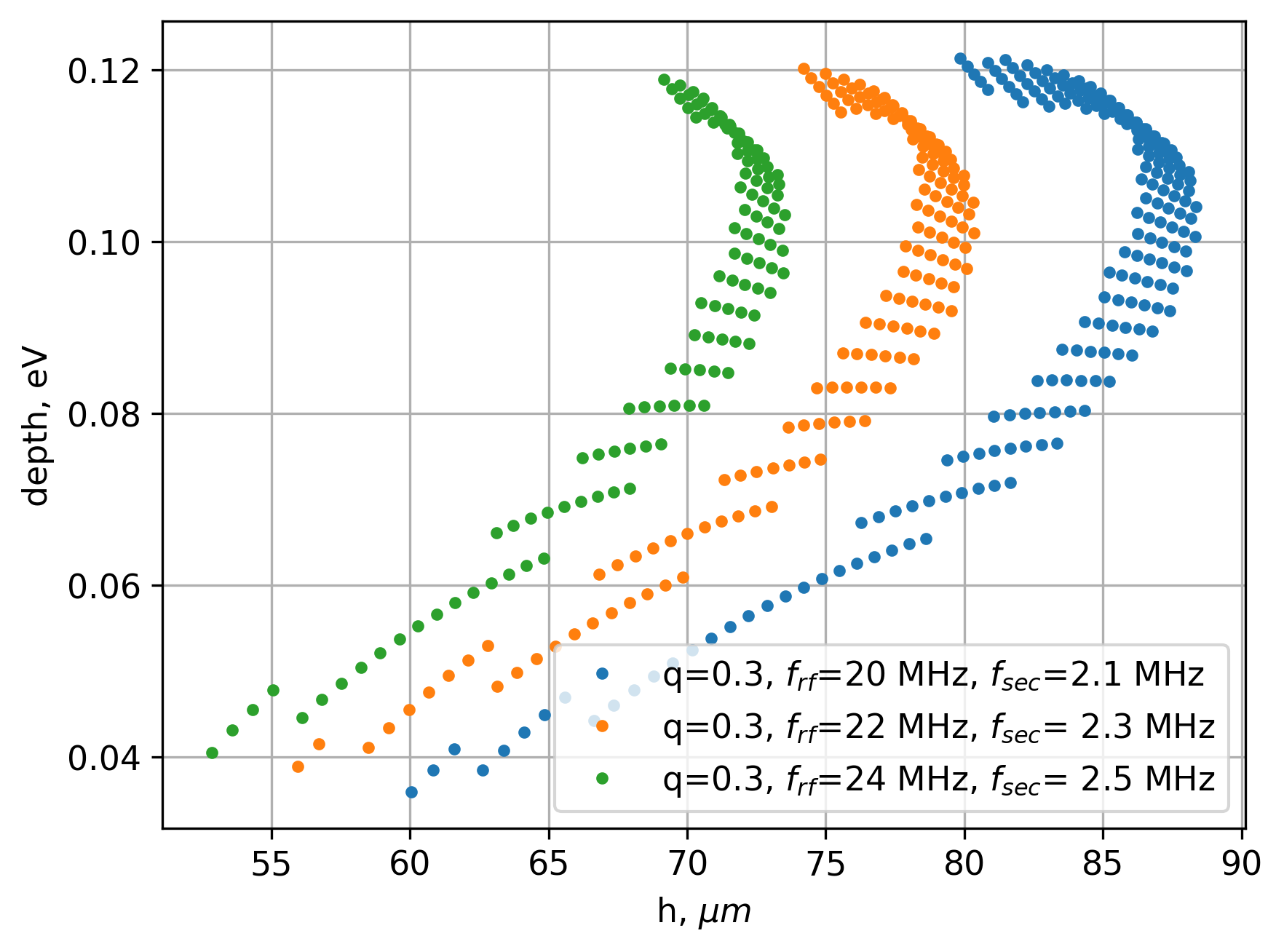}
    \caption{Fig. 5. Dependence of the trapping potential depth as a function of the distance $h$, $V_{rf}$ = 100 V. The scatter of points comes from a small variation of  $q = 0.300 \pm 0.008$.}
    \label{fig:depth}
\end{figure}

Next, we consider the potential depth. Fig. \ref{fig:depth} shows the dependence between it and the parameter $h$. Each group of points of the same color corresponds to a certain value of $f_{rf}$. The qualitative behavior can be described in the same way as in the previous paragraph. If $h$ is too large, no solutions exist. By lowering $h$, one value of the potential depth becomes possible. By further reducing $h$, we observe two achievable solutions, each corresponding to one of two intersections between lines with certain values of  $h$ and $f_{rf}$ given in Fig. \ref{fig:height} and \ref{fig:freq}.

By selecting the point with the maximum $h$ for any given $f_{rf}$, we achieve a potential depth of more than 0.1 eV, sufficient for ion trapping. As shown in Fig. \ref{fig:depth}, increasing the secular frequency reduces $h$, requiring a balanced solution. We determine the optimal values to be $f_{rf}=22\,$MHz and  $h \approx 80\,$$\mu$m. This region corresponds to $w_c \in [40, 60]\,$$\mu$m and $w_r \in [140, 200]\,$$\mu$m.


\section{Asymmetric trap}

The trap presented in Fig. 1 is symmetric, which means that both RF electrodes have the same width. In this case, two modes of the secular motion coincide with the $x$ and $y$ axes. This means that a cooling laser beam parallel to the trap surface will not effectively cool the motion perpendicular to the trap surface. To address this issue, an asymmetric trap with differing RF electrode widths can be employed. In such a configuration, both mode axes have a projection onto the laser beam, enabling effective cooling along both axes.

We denote the angle between the normal to the surface and the secular mode direction (closest to the vertical) as $\alpha$ and consider the design of an asymmetric trap. Assume that the RF electrodes have different widths: $w_{r}^{u}$ and $w_{r}^{d}$. The central electrode of width $w_c$ and outer DC electrodes remain grounded. To tilt the direction of secular modes, we vary $w_c$, $w_{r}^{u}$ in the boundaries defined by the previous analysis (this allows us to reduce the calculation grid) and $w_r^d$ in the range $[140, 500]\,$$\mu$m. The angle $\alpha$ is considered to be in the range from $10^{\circ}$  to $20^{\circ}$. On the one hand, $\alpha$ should be big enough to ensure effective cooling on both axes; on the other hand, better symmetry is usually more optimal in terms of trap parameters. The driving frequency and the field amplitude are fixed at $f_{rf} = 22\,$MHz and $V_{rf} = 100\,$V. The program computed the Hessian of the pseudopotential at its minimum. By diagonalizing the resulting matrix, we determined the directions of the vibrational axes. The resulting solutions were post-selected to satisfy the criteria  $h \approx$ 80 $\mu m$, $q \approx 0.3$, $\alpha \in [10, 20] ^{\circ}$.  Finally, the optimal solution was chosen that maximizes the trap depth. Its parameters are summarized in Table \ref{tab:geom}.

\begin{table}[h]
    \centering
    \begin{tabular}{ |c|c|c|c|c| } 
         \hline
         $w_c$ & $w_r^d$ & $w_r^u$ & $V_{rf}$ & $f_{rf}$\\ 
         \hline
         40 $\mu m$ & 160 $\mu m$ & 400 $\mu m$ & 100 V & 22 MHz\\
         \hline
         \hline
         $q$ & $h$ & $x_0$ & depth & $\alpha$\\ 
         \hline
         0.3 & 80 $\mu m$ & -11 $\mu m$ & 110 meV & 14$^{\circ}$\\
         \hline
    \end{tabular}
    \caption{Table 1. Optimal parameters of the asymmetric trap for the considered calculation grid. $x_0$ is the displacement of the ion across the $x$-axis caused by the trap asymmetry.}
    \label{tab:geom}
\end{table}

\begin{figure}[htp]
    \centering
    \includegraphics[width=7cm]{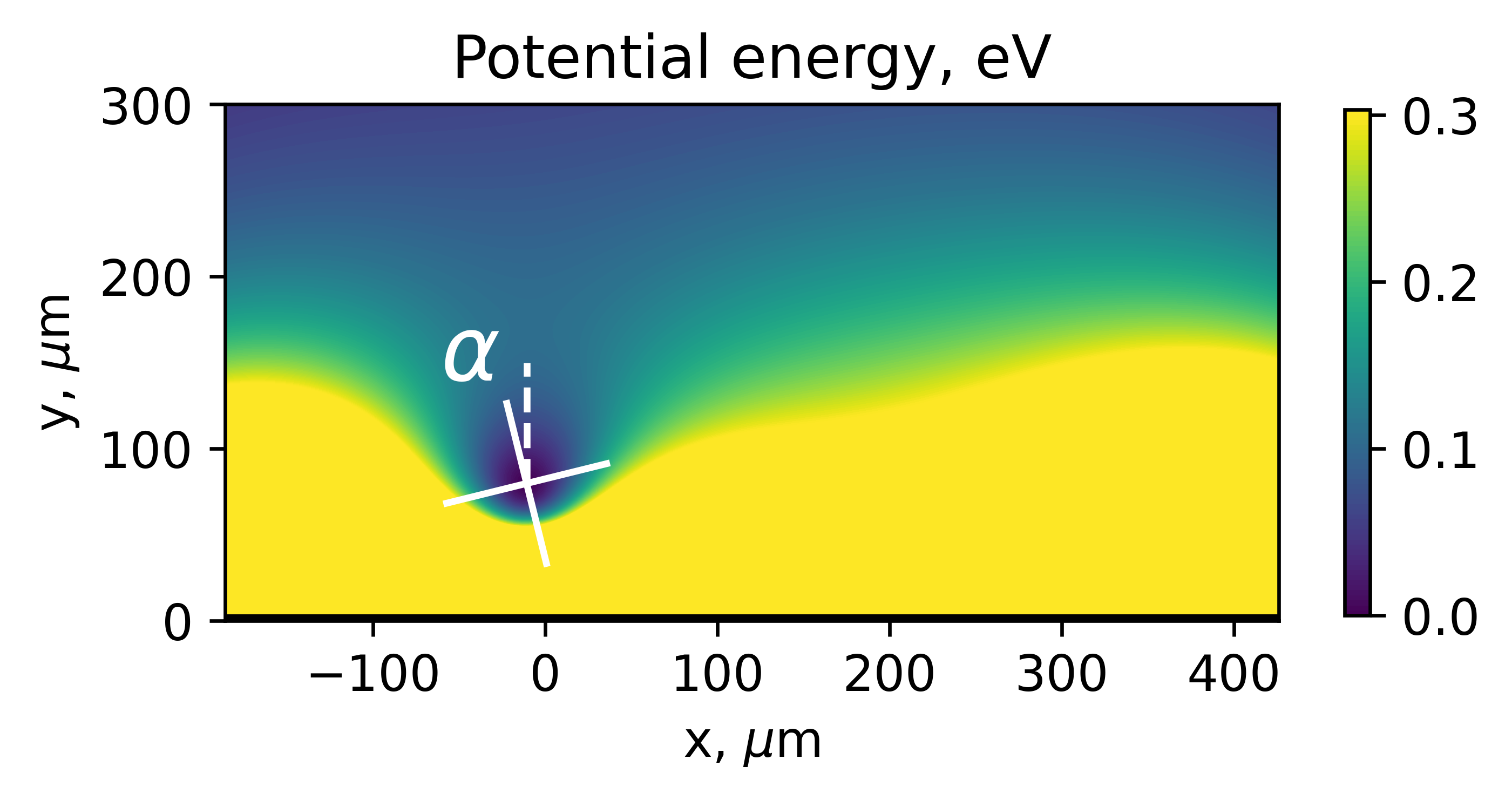}
    \caption{Fig. 6. Pseudopotential distribution above the trap surface in $x-y$ plane and axes of secular motion (white). The parameters of the trap are presented in Table \ref{tab:geom}.}
    \label{fig:potential}
\end{figure}

The corresponding pseudopotential distribution and principal axes of secular motion are presented in Fig. \ref{fig:potential}. The equilibrium position is above the $x-z$ plane at the height 80 $\mu m$ and is shifted by $x_0 = 11\:\mu m$ along the $x$-axis towards the thin RF electrode. The principal axes of secular motion are rotated by the angle of 14$^{\circ}$ towards this electrode.

\section{Ion chain separation}

The next step is to design the trapping potential along the $z$-axis. We simulate the lengths and positions of the outer electrodes that modify the axial DC potential, thereby defining the movement of particles across the z-axis. The separation of the ion chain into two parts, the basic operation of modern QCCD architecture, should be optimized to provide robust and reproducible operations. A chain of ions in an axial trapping potential can be separated by creating a potential barrier between neighboring ions (Fig.\ref{fig:chainsep3D}). The challenge in precisely dividing an ion chain lies in the fact that inter-ionic distances are typically less than 10 $\mu m$, while the distance to the electrodes is at least an order of magnitude greater. Under these conditions, it is impossible to create an electric potential peak narrow enough to resolve the distance between ions. Therefore, the most effective strategy for localizing chain separation is to minimize the width of the potential barrier.

\begin{figure}[htp]
    \centering
    \includegraphics[width=8cm]{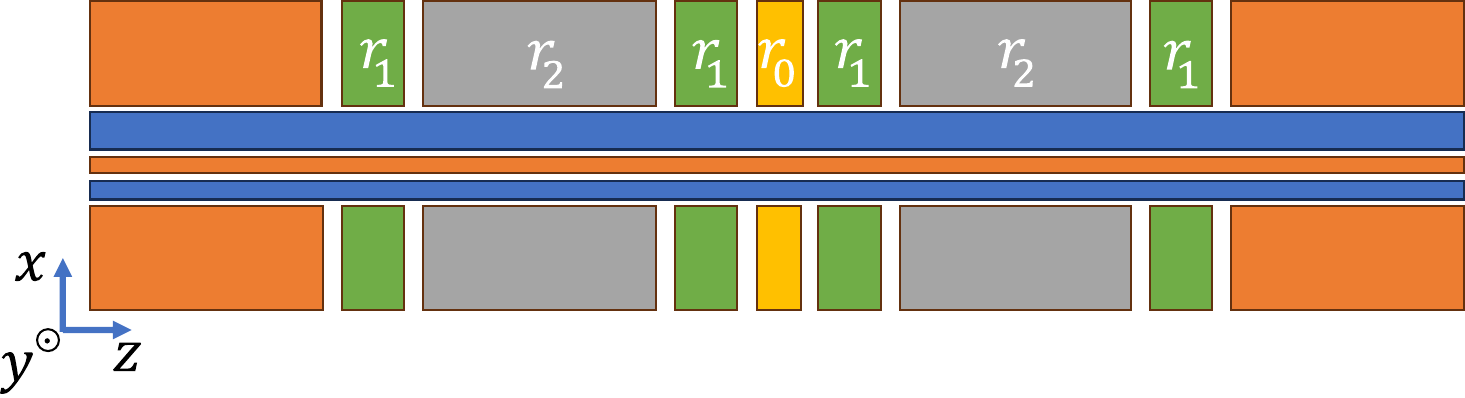}
    \caption{Fig. 7. Design of outer DC electrodes for ion chain separation. Here, $r_0$ is the length of the central separation electrode, $r_1$ is the length of the locking electrodes, and $r_2$ is the length of the middle electrode within each section.}
    \label{fig:dcelectrodes}
\end{figure}

Fig. \ref{fig:dcelectrodes} gives the suggested trap design. The trap consists of two sections, which comprise three pairs of electrodes: the central pair of  $r_2$ width (depicted as gray in Fig. \ref{fig:dcelectrodes}), and two side pairs (green) with the width of  $r_1$. The trap center has an electrode (colored yellow) with width $r_0$ to create a separating "wedge"\ potential barrier. We will consider the initial situation of a single ion chain confined in the very center of the trap by applying a positive potential to the pairs of grey electrodes and a negative potential to the pairs of central green and yellow electrodes. At the end of the splitting procedure, each of the two sections confines its separate ion chain. Here we will consider the optimal geometry of the electrodes involved in the process of chain splitting. The dynamics of ions during the separation procedure and the recapture process of sub-chains into the corresponding trap zones are beyond the scope of this work.

\begin{figure}[htp]
    \centering
    \includegraphics[width=8cm]{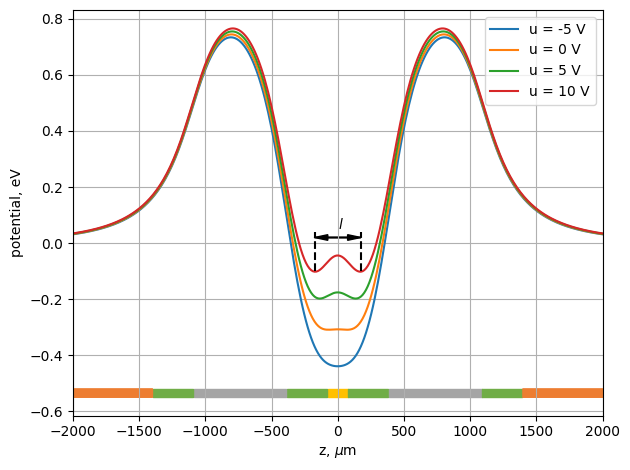}
    \caption{Fig. 8. Trap potential along the axial direction. The configuration of RF electrodes is presented in Table \ref{tab:geom}, and the configuration of DC electrodes is presented in Table \ref{tab:dcparams}. Colored rectangles at the bottom represent the lengths and z-axis positions of outer electrodes; colors correspond to Fig. \ref{fig:dcelectrodes}.}
    \label{fig:dcpotential}
\end{figure}

At the start of the splitting procedure $10\,$V is applied to the central (grey) electrodes and $-10\,$V to the side (green) electrodes near the separation electrode; all outer side electrodes are grounded. Then one increases the voltage $u$ on the separation electrode (yellow) from -5 V to 10 V to raise the barrier and create a double-well structure. The evolution of the resulting axial potential is shown in Fig. \ref{fig:dcpotential}.
Voltages in the range of $10$ V are easily achievable in a laboratory without getting much voltage noise. 

To ensure effective separation, the slope of the potential barrier between the split ion chains should be maximized. Assuming a fixed depth for both wells after splitting, this condition can be met by minimizing the distance between the wells' minima (denoted as $l$).
To determine the optimal geometrical configuration for the smallest possible distance, we fix the depth of each potential well after separation to $0.05\,$eV and vary the widths of the electrodes within the following ranges: $r_0 \in [50, 200]\,$ $\mu$m, $r_1 \in [100, 300]\,$ $\mu$m, $r_2 \in [500, 800]\,\ \mu$m. The parameters of the optimal configuration, yielding the distance of 350 $\mu$m between the wells' minima, are presented in Table \ref{tab:dcparams}.

\begin{table}[h]
    \centering
    \begin{tabular}{ |c|c|c|c| } 
         \hline
         $r_0$ & $r_1$ & $r_2$ & $l$\\ 
         \hline
         150 $\mu m$ & 300 $\mu m$ & 700 $\mu m$ & 350 $\mu m$\\
         \hline
    \end{tabular}
    \caption{Table 2. Parameters of optimal configuration for ion chain splitting with trap presented on Fig. \ref{fig:dcelectrodes}.}
    \label{tab:dcparams}
\end{table}

The proposed design of the surface Paul trap splitting element is easy to manufacture and allows to create a double-well potential in the z-direction of the trap geometry. By optimizing the lengths of the electrodes in the axial direction, we ensured the maximum slope of the central potential barrier.

\section{Conclusion}


In this study, we detailed the design process of a surface Paul trap optimized for confining and manipulating ytterbium ions. We began by calculating the distance from the ion to the trap surface and the potential depth, establishing a configuration that holds the ion at a height of $h = 80$ $\mu$m and can trap ions with energies below $110$ meV. During optimization, we considered the stability parameter and the secular frequency of the trap, ensuring their values remained close to those proven effective in previous experiments. Subsequently, we adjusted the principal axes of secular motion by an angle of $\alpha = 14^{\circ}$ by introducing asymmetry in the RF electrodes, achieving conditions for efficient laser cooling. Additionally, we equipped the trap with a feature for ion chain division using the outer DC electrodes to create a well-localized separating potential barrier. Moving forward, we aim to implement the calculated design on a microchip and conduct ion-trapping experiments.

\section*{Funding}

This work is supported by RSF grant \textnumero\,24-12-00415.

\section*{Conflict of interest}

The authors of this work declare that they have no conflicts of interest.

\bibliographystyle{IEEEtran}
\bibliography{bib}

\end{document}